\documentclass{article}

\usepackage{arxiv}

\usepackage{array}
\usepackage[utf8]{inputenc} 
\usepackage[T1]{fontenc}    
\usepackage{hyperref}       
\usepackage{url}            
\usepackage{booktabs}       
\usepackage{amsfonts}       
\usepackage{nicefrac}       
\usepackage{microtype}      
\usepackage{amsmath}
\usepackage{lipsum}		
\usepackage{graphicx}

\usepackage{subcaption}
\usepackage[sorting=none,style=ieee, backend=biber]{biblatex}
\usepackage{doi}

\title{Improving Interface Physics Understanding in High-Frequency Cryogenic Normal Conducting Cavities}

\author{ \href{https://orcid.org/0000-0001-6011-3435}{\includegraphics[scale=0.06]{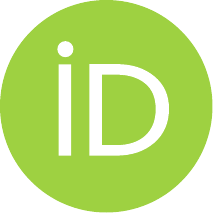}\hspace{1mm}Gerard E.~Lawler}\thanks{\url{https://gelawler.github.io/}}\\
	Department of Physics \& Astronomy\\
	University of California, Los Angeles\\
	Los Angeles, CA 90045 \\
	\texttt{gelawler@physics.ucla.edu} \\
	\And
	Fabio Bosco \\
	Department of Physics \& Astronomy\\
	University of California, Los Angeles\\
	Los Angeles, CA 90045 \\	\And
	James B.~Rosenzweig \\
	Department of Physics \& Astronomy\\
	University of California, Los Angeles\\
	Los Angeles, CA 90045 \\
}

\renewcommand{\shorttitle}{\textit{arXiv} Template}

\hypersetup{
pdftitle={A template for the arxiv style},
pdfsubject={q-bio.NC, q-bio.QM},
pdfauthor={David S.~Hippocampus, Elias D.~Striatum},
pdfkeywords={First keyword, Second keyword, More},
}

\begin{document}
\maketitle

\begin{abstract}
	As progress towards real implementations of cryogenic high gradient normal conducting accelerating cavities continues, a more mature understanding of the surface physics in this novel environment becomes increasingly necessary. To this end, we here focus on developing a deeper understanding of one cavity figure of merit, the radiofrequency (RF) surface resistivity, $R_s$. A combination of experimental measurements and theory development form the basis of this work. For many cases, existing theory is sufficient but there are nuances leading to systemic errors in prediction which we address here. In addition, for certain cases there exist unexpected local minimum in $R_s$ found at temperatures above 0K. We compare here several alternative models for RF surface resistivity those which incorporate thin film like behavior which we use to predict the location of the local minimum in surface resistivity. Our experimental results focus on C-band frequencies for the benefit of several future cryogenic linear accelerator concepts intended to operate in this regime. To this end we have measured factor of $2.89\pm 0.05$ improvements in quality factor at $77$K and $4.61\pm 0.05$ at 45K. We further describe the test setup and cooling capabilities to address systematic issues associated with the measurements as well as a comparison of RF cavity preparation and the significant effect on $R_s$. Some implications of our measurements to linear accelerators combined with the theoretical considerations are extended to a wider ranger of frequencies especially the two additional aforementioned bands. Additional possible implications for condensed matter physics studies are mentioned. 
 
\end{abstract}


\section{Introduction}
An area of significant interest in particle accelerator research is the development of accelerating cavities which can support very high field gradients. Field gradients are usually limited by breakdown rates (BDR) so significant experimental progress has been made in the advancement of technologies which reduce BDR thus allowing for higher field gradient \cite{SIMAKOV2018221, app131910849}. In particular, empirical studies have observed ultra high peak fields in excess of 500 MV/m at 45K which correspond to 250 MV/m accelerating fields \cite{Cahill-2017}. Significant consideration has been given to X-band $\left(8-12 \text{ GHz}\right)$ and S-band $\left(2-4 \text{ GHz}\right)$ for a number of reasons \cite{PhysRevAccelBeams.24.081002, Cahill:IPAC2016-MOPMW038,Rosenzweig-2018,Rosenzweig-2019}. C-band $\left(4-8 \text{ GHz}\right)$ however presents a number of advantageous features especially from the stand point of a practical achievable linear accelerator based on high gradient normal conducting cryogenic cavities. Several future linear accelerator concepts have been proposed which in this frequency range which utilize the high gradients made possible from cryogenic operation \cite{UCXFEL,bai2021c,lawler2021rf}. 


Theoretical progress regarding surface behavior especially in the context of BDR has been slow largely owing to complicated nature of the phenomena involving extreme environments on varying time and length scales \cite{cahill2017ultra}. The gap in understanding is in several ways insufficient for the needs of a growing field of research into high gradient cavities. Indeed as progress towards cryogenic operation of normal conducting cavities continues, a more mature understanding of the surface physics is necessary. This is especially true at cryogenic temperatures in an ultra-high gradient regime. Still there is a significant body of work to which we can refer for our specific RF oscillating field case, especially developed in the context of DC breakdown experiments and theory \cite{PhysRevSTAB.12.102001, PhysRevSTAB.15.071002}. The theory here is complex and involved since it involves multiple length and time scales. 

To this end we want to develop improvements which are more modest in scope and simple to compute but still notabley improve understanding especially for experimental realization. Specifically, we here focus on one important figure of merit (FoM), the radiofrequency (RF) surface resistivity $R_s$. We attempt to iteratively improve predictions and understanding of this single FoM from which we can derive more accurate experimentally relevant near term useful predictions. The relevance of $R_s$ as a proxy for general cryogenic cavity performance is multidimensional but can be explained primarily with reference to the RF pulse heating which oscillating field cavities experience per cycle. Through the affect on pulse heating, $R_s$ then becomes useful in informing BDR behavior \cite{10.1063/5.0084266}. These numbers have been analytically computed in the past to calculate for example temperature rise in room temperature cavities \cite{pritzkau2001rf}. Extension to cryogenic temperature then becomes possible as long as one has an  understanding of low temperature $R_s$. 

In the high temperature limit, computing $R_s$ is as simple as using Maxwell's equations with appropriate boundary conditions to compute the real component of the complex impedance as shown in Equ. \ref{eqn:RS}.

\begin{equation}
\label{eqn:RS}
    R_s\left(T\rightarrow \infty\right) = Re\left(Z_s\right) = \sqrt{\frac{2\pi f \mu_0\rho}{2}}
\end{equation}

From this expression we observe the implicit dependence on temperature via the bulk electrical resistivity $\rho$ and to a lesser extant the eigenmode frequency $f$. The simplest theory for temperature dependence of $\rho$ which we have comes via the work of Bloch and Gruneisen \cite{RevModPhys.62.645, nasr2021distributed}

\begin{equation}
\label{eqn:BG}
\rho\left(T\right)=A\left(\frac{T}{\Theta_R}\right)^n\int_0^{\Theta_R/T}\frac{t^n}{\left(e^t-1\right)\left(1-e^{-t}\right)}dt+C
\end{equation}

\begin{figure}[!htb]
   \centering
   \includegraphics*[width=1\columnwidth]{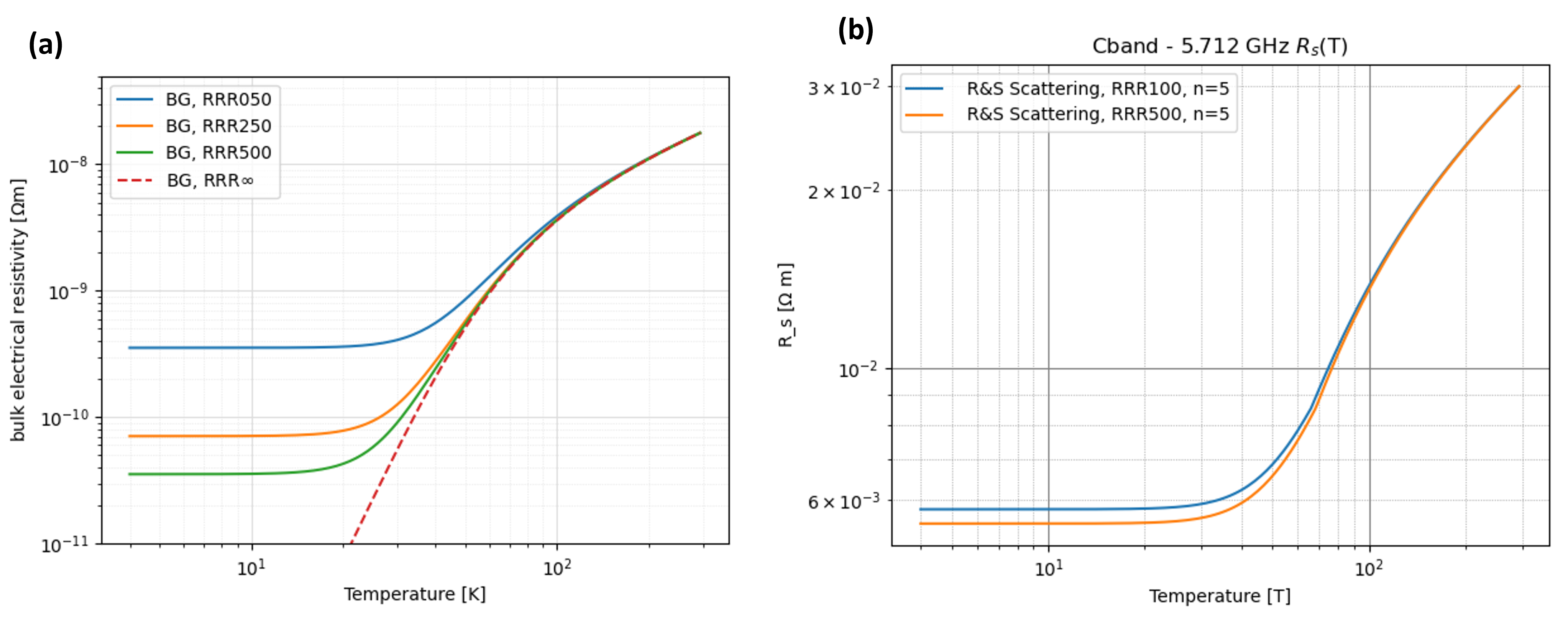}
   \caption{\textbf{(a)} Bloch-Gruneisen temperature dependent bulk resistivity from Equ. \ref{eqn:BG} with $n=5$ \textbf{(b)} Surface resistivity for Cband cavity using Reuter-Sondheimer-Chambers theory to compute limits and using patching formula.}
   \label{fig:BGn5}
\end{figure}

Where $A$ is a scaling constant, $\Theta_R$ is the Fermi temperature and $C$ is an constant left as a free parameter to adjust for the residual resistivity ratio (RRR) defined as the ratio of room temperature resistivity to that at 4K. The remaining paremeter $n$ is dimensionless and established by the dominant scattering mechanism at play for the situation. The value used in the past has been $n=5$ which is true for ideal metals \cite{nasr2021distributed, 10.1063/1.555614}. We can compute this temperature dependent resistivity and plot for the case of a number of different RRR values. These are showed in Fig. \ref{fig:BGn5}. There are several features here to note, namely that for $C=0$, the resistivity drops to 0 at 0K such that the $RRR\rightarrow \infty$.

In the low temperature limit one common theory used for calculation of $R_s$ is derived from the work of Reuter and Sondheimer \cite{reuter1948theory} with further significant contribution from Chambers \cite{chambers1952anomalous}. The theory was formulated to explain the systematic underestimate of $R_s$ as the surface approaches 0K, a phenomena termed the anomalous skin effect (ASE). The primary effect here is the differing functional dependence on temperature of the electron mean free path length (MFPL) and RF skin depth, $\delta$. MFPL reduces much faster than $\delta$ and cryogenic temperatures such that at cryogenic temperatures MFPL becomes large compared to $\delta$, more scattering events occur and $R_s$ is reduced. It is common to compute the asymptotic value of surface resistivity at $0$ K. As shown in Equ. \ref{equ:RS_0K}. It is worth noting for future discussion that there is no explicit dependence on bulk properties of the metal in this theory. 

\begin{equation}
    R_s\left(T\rightarrow 0\right) = Z_0\left[\frac{\sqrt{3}v_f}{16\pi c}\left(\frac{\omega}{\omega_p}\right)^2\right]^{\frac{1}{3}}
    \label{equ:RS_0K}
\end{equation}

A patching function is then used to interpolate between the two regimes with dimensional parameters $a$ and $b$ given by the relative proportion of diffusive to specular scattering of electrons in the surface material as shown in Equ. \ref{equ:patching_eq}.

\begin{equation}
\label{equ:patching_eq}
    R_s\left(T\rightarrow 0\right)=R_{\infty}\left(1+a\alpha^{-b}\right)\qquad \text{for } \alpha\ge 3
\end{equation}

The dimensionless parameters range of order unity and different values are used by different analysis to reflect once again the relavent scattering behavior in the metal \cite{chou1995anomalous}. In the case of Reuter and Sondheimer's original formulation these values correspond to $a=1.004$ and $b=0.3333$; values which were are circumstances experimentally established by Chambers. Of note then is that predictions are useful but not sufficient for assessing $R_s$ from first principles in the regime of practical consideration to normal conducting cryogenic RF cavities. Intended temperature ranges for our normal conducting cavities are primarily in the range of $35-77$ K, unfortunately placed directly in the interpolation regime. 

In addition, more advanced specially designed \emph{mushroom dome} RF cavity measurements deviate from the aformentioned model in that there exists a local minimum in $R_s$ around $30$ K \cite{Cahill-2017}. Furthermore, subtle bulk material property dependence was observed \cite{cahill2017ultra}. Neither of these features are predicted at all with the existing framework of calculation. Observing that the complex nature of the theory of BDR would be difficult to incorporate into multiphysics simulations of the type found in CST or HFSS, we are thus we are justified in seeking simple improvements of $R_s$ predictions.

In the field of RF cavities, $R_s$ is measured often measured via the unloaded quality factor $Q_0$. $R_s$ for these simple cavities are then derived from $Q_0$ as an inverse quantity scaled by some geometric factor $G$. More information on the systemics this introduces into our measurements and simulations will be covered in the next section but here we note that this introduces the necessity to precisely consider another bulk property in addition to the bulk resistivity, the coefficient of thermal expansion. Now there is the additional implicit temperature dependence via the changing eigenmode $\omega$ at which we measure $S11$ and the changing cavity coupling via the RF antenna. More on the temperature dependent COE values will be explained in the methodology section.

\section{Methodology}

Our intention is thus to compute with full transparency, the RSC predictions for a number of different scattering parameters make adjustments along the way where necessary. We will further introduce more subtle thin film physics effects in RF cavities and propose simple alternative models. An experimental setup and measurement suite was designed for measuring $R_s$ in Cband RF cavities in order to validate the models down to at least application relevant temperatures. 

\subsection{Theory}

\begin{figure}[!htb]
   \centering
   \includegraphics*[width=\columnwidth]{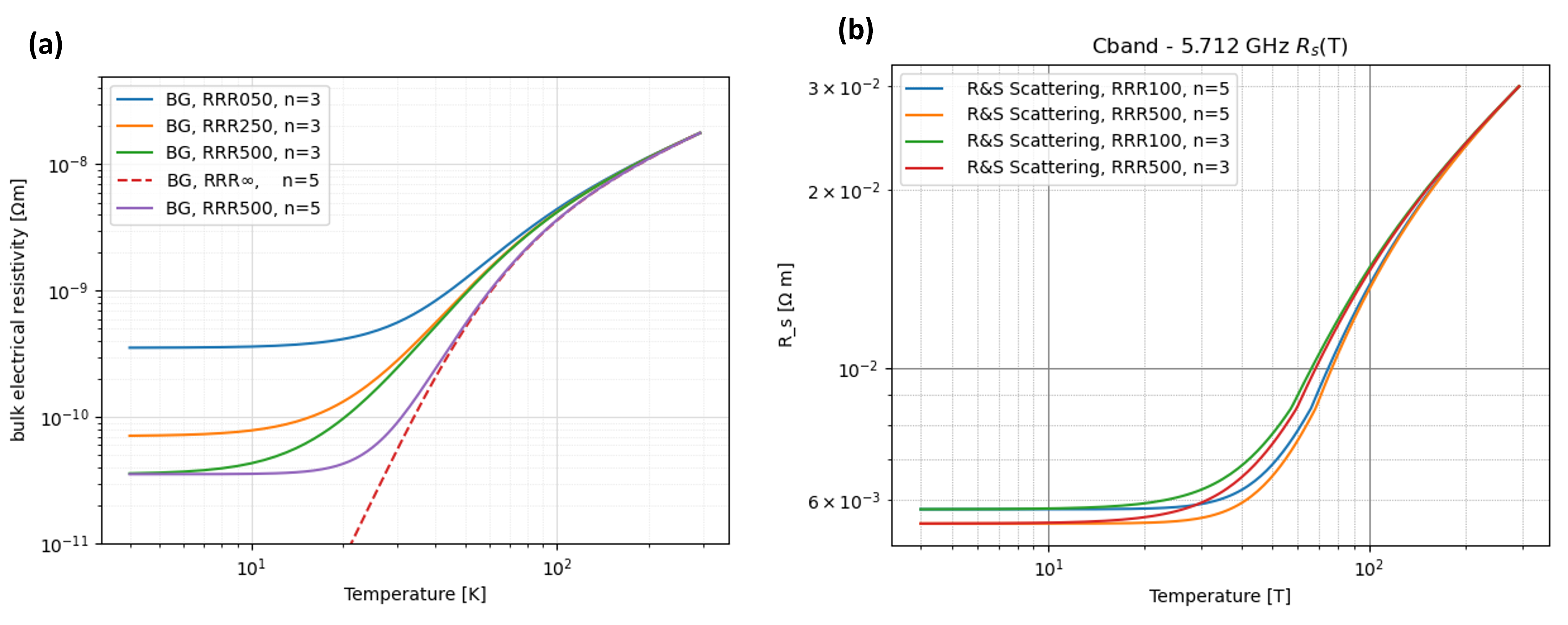}
   \caption{Bloch-Gruneisen temperature dependent bulk resistivity with $n=3$ with this exception of $C_{BG}=0$ and a RRR500 curve which still use $n=5$ for comparison. The feature to note here is the significant deviation at intermediate values.}
   \label{fig:BGn4}
\end{figure}

Our first modification beyond the introduced theory $R_s$ comes from a nuance regarding the BG equation itself. We note that the value of $n=5$ is only strictly speaking true in certain circumstances special circumstances \cite{10.1063/1.555614}. For transition metals, like copper, a more accurate value is $n=3$. The calculations with this improvement are shown in Fig. \ref{fig:BGn4}. The notable significance here is that the asymptotic high temperature and low temperature values are agnostic to this change but the intermediary values decidedely are not. For a not unreasonable number of $RRR=500$, the 77K $R_s$ is increased by a factor of $\approx 1.3$ and nearly $2$ at $45$K.

\subsubsection{Thin Film Physics} To address the existence of experimental local minimum at intermediate temperatures we want to explore more in depth the space of interface physics and in doing so indeed find that there are situations where $R_s$ does not monotonically decrease with temperature. Indeed for certain circumstances it increases with decreasing temperatures within the intermediate range of temperatures with which we are concerned. This is called the Gurzhi effect and is well established for thin film physics representing a novel regime where electrons in the material take on an effective viscosity \cite{gurzhi1963minimum,doi:10.1143/JPSJ.25.390, Polini_2020}. The justification of considering one skin depth as an effective thin film with thickness $\delta$ has been made and we reiterate it here \cite{gurzhi1964contribution}. To this end two alternative models we will be our focus here: the qualitative components of the viscous electron model developed by Gurzhi and a more simple application of thin film resistivity developed by Fuchs \& Sondheimer \cite{pichard1985alternative}. Both predict local minimum in $R_s$. Based on these improved models for surface resistivity we find a nonzero optimal temperature for normal conducting copper cavities as a function of RRR.

\begin{figure}
    \centering
    \begin{subfigure}{.48\columnwidth}
         \centering
   \includegraphics*[width=0.85\columnwidth]{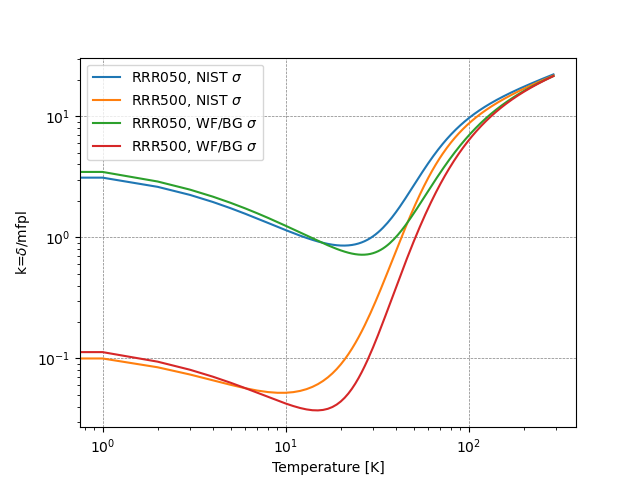}
   \caption{Ratio of RF skin depth (calculated classically) to $\ell_{eff}$ calculated via NIST log functions}
     \end{subfigure}
     \hfill
     \begin{subfigure}{.48\columnwidth}
         \centering
   \includegraphics*[width=0.85\columnwidth]{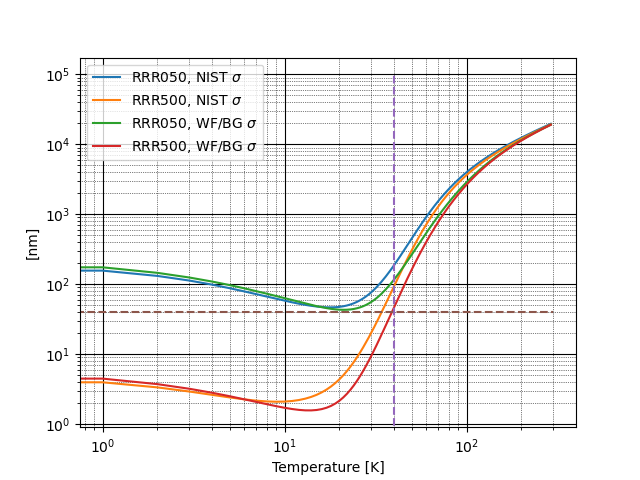}
   \caption{$\delta$ times the ratio from plot \textbf{(a)} showing an idea of where the Gurzhi effect may occur in cavities relevant to Cband cryogenic operation}
     \end{subfigure}
     \caption{Length scales involved in theoretical computations of $R_s$ relevant to our C-band cavity dimensions}
    \label{fig:Length_scales}
\end{figure}

As an application of Gurzhi theory, we first must consider some of the length scales involved in computation of $R_s$ for our C-band cavities. We plot these in Fig. \ref{fig:Length_scales}. Specifically we have the same asymptotic values as as existing theory with exception of an intermediary regime where as temperature increases, the resistance increase. This regime corresponds to the situation where the effective mean free path length is approximately equivalent to the following expression. 

\[\ell_{eff}\sim \frac{\delta^2}{\ell_{ee}}\]

That is to say the effective electron mean free path becomes approximately equal to the square of the sample size divided by the electron electron mean free path length. For our effective mean free path length we can use NIST polylog fit curves published from measured data which relate the thermal conductivity to temperature dependence. We can then relate this to the $\ell_{eff}$ with which it should scale linearly. We further consider Mathiessen’s rule such that we can say the following \cite{calatroni2020materials}

\[\ell_{eff}=\left(\frac{1}{\ell_{phonon}}+\frac{1}{\ell_{impurities}}+\cdots\right)^{-1}\]

So for the case where $\ell_{ee}$ is largely a material dependent property and thus independent of temperature we can compute the regime where this behavior should occur for the case of $RRR450$ where if $\ell_{ee}\approx 40$ nm \cite{10.1063/1.4942216} we would have the Gurzhi effect regime kicking in around $35-40$K. In addition, for a low purity material like $RRR050$ we would not expect the regime to be present at all. We can see this in Fig. \ref{fig:Length_scales}. Our intention thus is to cool to below this value for our existing pillbox structures to determine this negative slope $dR_s/dT$ in experimental data. 

A fully analytic application of Gurzhi theory is too involved for the time being so considering the need for a more computationally tractable theory. For simplicity and the sake of computational feasability, we also propose a model where since the RF skin depth is treated as a thin film sample with thickness equal to the skin depth we calculate directly the resistivity of a thin film. We refer here to the formulation of the Fuchs and Sondheimer theory to compute an thin film resistance\cite{pichard1985alternative}. With constants of proportionality composed of elementary constants and material properties not explicitly dependent on temperature. So we have the following where the value $p$ relates the proportion of specular to diffusive scattering at the boundary

\[\frac{\rho_{film}}{\rho_{bulk}} \approx\left[1.0-\frac{C_0}{\rho^{3/2}}\left(1-p\right)\right]^{-1} \]

We again make the assumption that the impurity mean free path length dominates at low temperature and is independent of temperature. For this model we assume the case of ideal metals, not transition metals, so we will limit our calculations in these cases to $n=5$. Also since $p$ is free parameter here we can make a semi-empirical model where this is scaled to previous data. This can then be said to be a near total specular scattering case since such that $1-p$ is vanishingly small. We can then use this thin film resistivity in place of the bulk resistivity in our existing formalism for computing $R_s$. Plotting these results as a function of temperature gives is shown in Fig. \ref{fig:Lawler_phase_space}


\begin{figure}[!htb]
   \centering
   \includegraphics*[width=0.5\columnwidth]{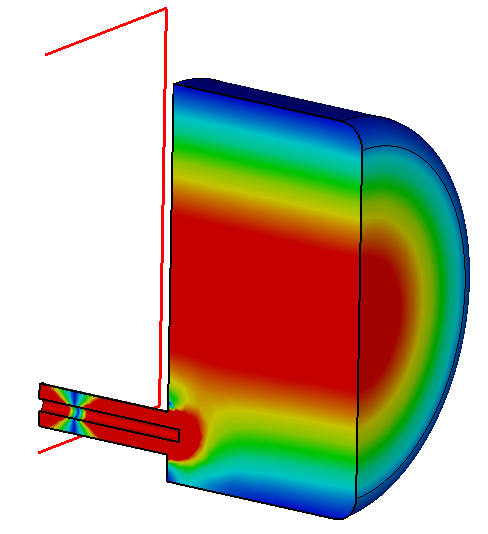}
   \caption{E-field profile in Cband cavity with antenna port}
   \label{fig:cavity_fields}
\end{figure} 

\subsection{Experimental Methods}

As mentioned in the introduction, experimentally the easiest method to measure the $R_s$ in RF cavities is via unloaded quality factors, $Q_0$. In order to do this we insert microwave antennas into the cavities and measure a the reflection coefficient called $S11$ produced from a low level input signal. $Q_0$ can then be computed from the shorted position as the central eigenmode frequency divided by the $f3db$ bandwidth both of which are measured on a vector network analyzer (VNA). We are then left with the need to calculate the geometric factor $G$ which related $Q_0$ to $R_s$ in the following manner.

\begin{equation}
    R_s=\frac{G}{Q_0}
\end{equation}

The resonant cavities that we use are as close to ideal cylindrical resonant cavity as possible, a geometry commonly called a pillbox. Such a cavity is ideal since it has analytic solutions for the eigenmodes in terms of cylindrical Bessel functions from which we can quickly compute analytic curves based on direct material properties. These offer useful checks of measured data during experiment in addition to full multiphysics simulations which we later perform in CST. The idealized Bessel function fields are depicted in Fig. \ref{fig:cavity_fields}.

The RF geometry of the cavities are machined as close to the idealization as possible. There are of course perturbations from the placement of four features in particular: an RF antenna, a vacuum port, a fillet for machining considerations, and the possibility of excess braze material in the edge where the cavity cap is added. These features are replicated in the CST multiphysics model for simulation. The consideration is now especially notable due to our attempt to remove as many systematics as possible in a measurements since high precision is needed. 

There is an additional subtlety associated with $G$ also since it has an implicit dependence on temperature via the coefficient of thermal expansion we can consider it separately in simulation. We can analytically solve for the $G$ for the case of the ideal pillbox cavity to obtain the following expression relating $Q_0$ and $R_s$

\begin{equation}
    R_s=\frac{\chi_{01}\eta}{2}\frac{1}{1+a/h}\frac{1}{Q_0}
\end{equation}

Where $a$ and $h$ are the radius and height of the cylinder. Thankfully then for an isotropic material the effect contraction cancels out and $G$ remains temperature independent. For completeness we simulate the situation with the real geometry with the expectation that the additional effect of antenna cooling and contraction should be of vanishing significance if any measurable at all. 

\begin{figure}[!htb]
   \centering
   \includegraphics*[width=\columnwidth]{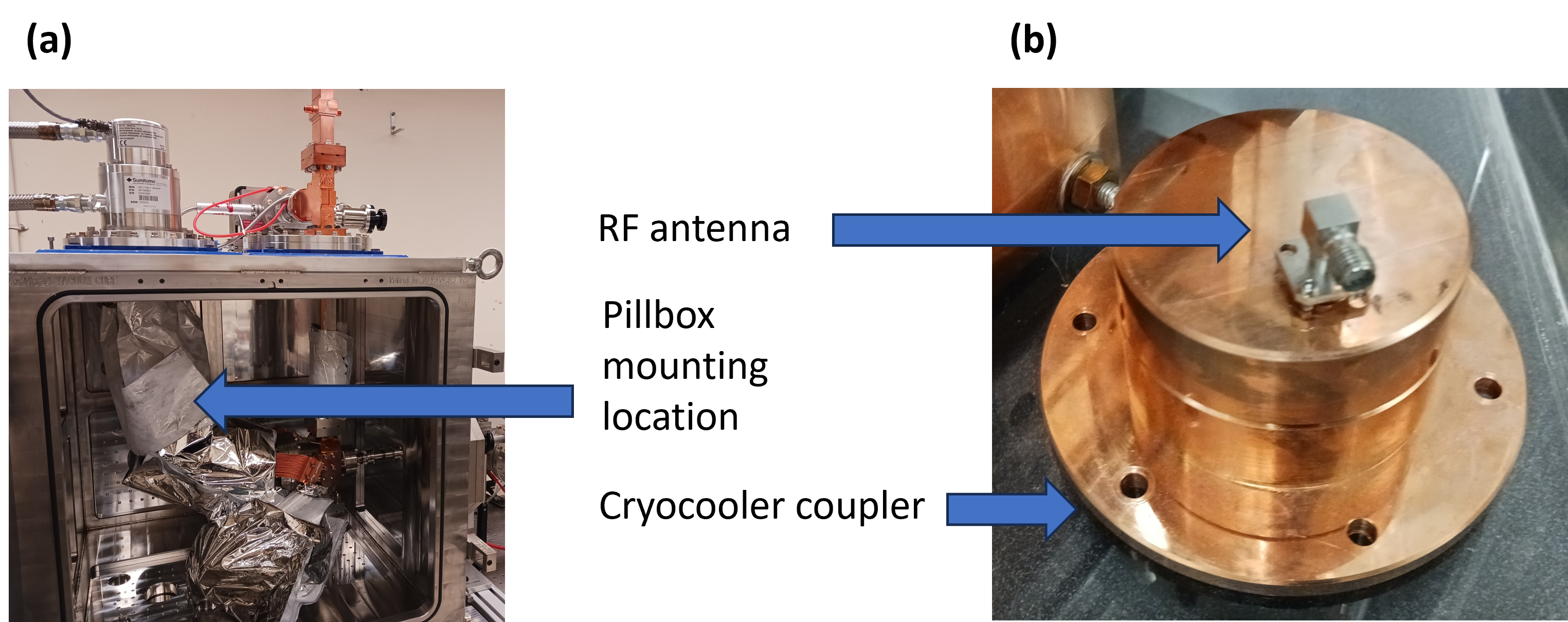}
   \caption{\textbf{(a)} CYBORG beamline cryostat which can be configured for LLRF tests. Including measurements presented here. \textbf{(b)} COMEB manufactured Cband cavity measured within CYBORG cryostat to measure data here presented.}
   \label{fig:CYBORG_setup}
\end{figure} 

With respect to the cryogenics we have commissioned at UCLA multiple new cryocoolers for novel RF testing including a single stage low temperatuer high power cryocooler used for a new CrYogenic Brightness Optimized Radiofrequency Photogun (CYBORG) \cite{lawler:ipac2022-thpost046}. In parallel to development of the photogun we have the versatility to configure our cryostat for pillbox cavity testing by direct coupling to the cooler heat exchanger. The CYBORG cryosat openned for access is shown in Figure \ref{fig:CYBORG_setup} along with the cavity under testing. The cooler relevant to these measurements is a Sumitomo Heavy Industries CH-110LT. Because of the high cooling capacity of this device our setup simply consists of the pillbox cavities mounted to the cold head via a simple polished C101 copper coupling plate. The composite system of cooler, coupler, and cavity are then wrapped in 14 layers of multi-layer insulation (MLI) with small penetration for temperature sensors and the RF antenna cabler. For cooling, the device under test is then placed within the CYBORG vacuum cryostat and pumped to around $10^{-5}$ torr.

\section{Results}

\begin{figure}
    \centering
    \begin{subfigure}{.49\columnwidth}
         \centering
   \includegraphics*[width=0.85\columnwidth]{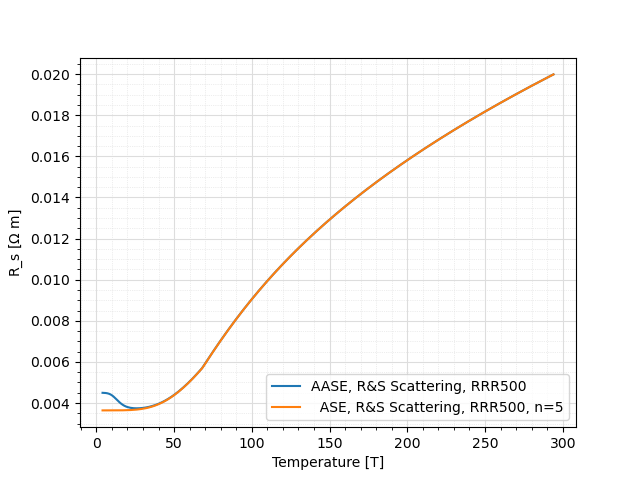}
   \caption{Single curve from contour plot in \textbf{(b)} shown compared to existing Reuter-Sondheimer-Chambers formulation}
     \end{subfigure}
     \hfill
     \begin{subfigure}{.49\columnwidth}
         \centering
   \includegraphics*[width=0.85\columnwidth]{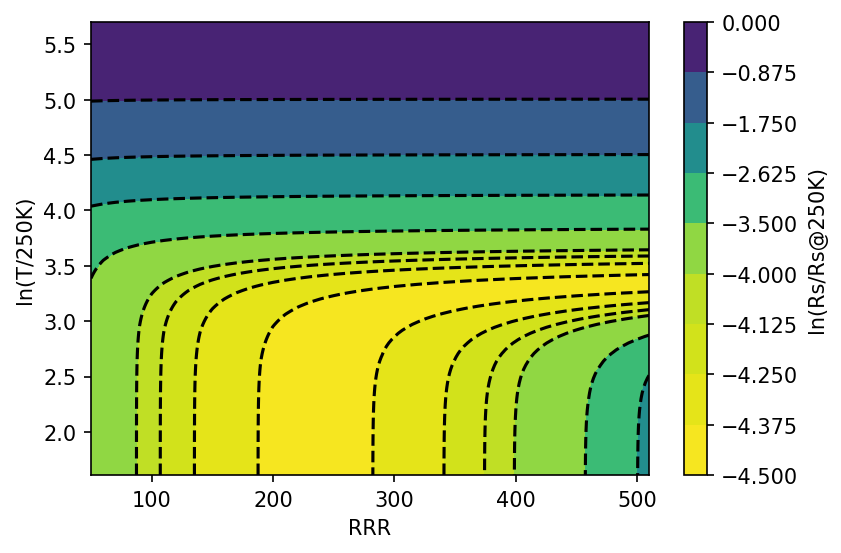}
   \caption{Solution space of predictions for various RRR values as a function of temperature.}
     \end{subfigure}
     \caption{Results of simplified Fuchs-Sondheimer style toy model used to compute $R_s$(T) in our relevant regime of Cband frequency.}
    \label{fig:Lawler_phase_space}
\end{figure}

We first address the results of our examination of improved simple models. For the Fuchs and Sondheimer case we note now they because of the additional dependence of bulk resistivity there is now additional dependence on RRR, a bulk material property. The theory then also predicts a local minimum around 25K for the high purity case. For a full space of solutions over many values of RRR we calculate a sweep for the contour plot shown in Fig. \ref{fig:Lawler_phase_space}. 

\begin{figure}
    \centering
    \begin{subfigure}{.49\columnwidth}
         \centering
   \includegraphics*[width=0.85\columnwidth]{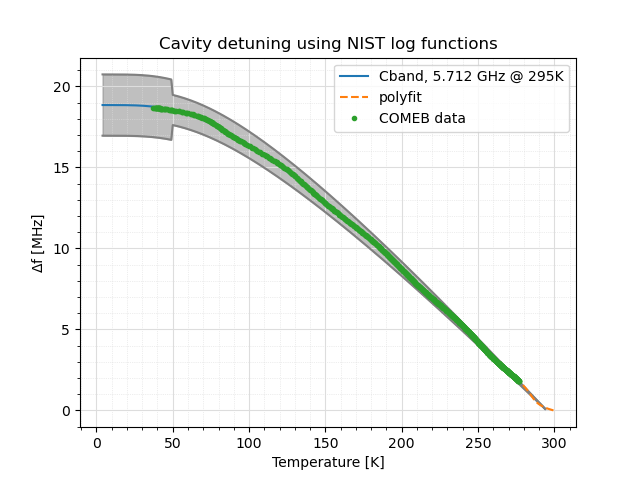}
   \caption{Cavity detuning as a function of temperature with uncertainty derived from published NIST curved for coefficient of thermal expansion as function of temperature.}
     \end{subfigure}
     \hfill
     \begin{subfigure}{.49\columnwidth}
         \centering
   \includegraphics*[width=0.85\columnwidth]{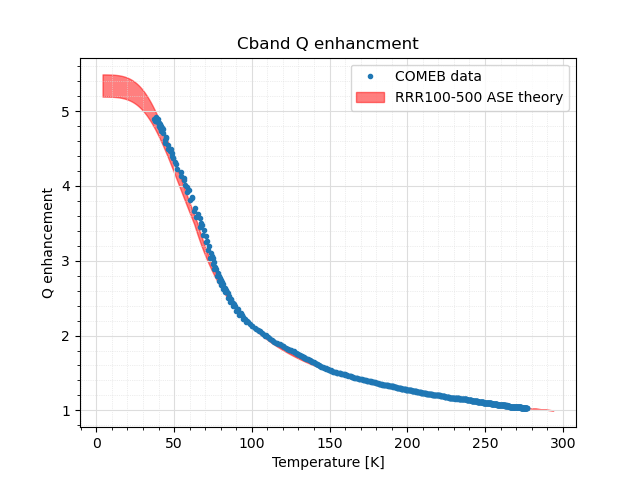}
   \caption{Q factor enhancement measured in-situ via S11 curve fitting algorithm}
     \end{subfigure}
     \caption{Data from measured COMEB pillbox cavity during cryogenic testing. Curves represent warming of cavities after cool down to minimum value of 38K}
    \label{fig:pillbox_data}
\end{figure}

Contained within Fig. \ref{fig:pillbox_data} are some of the results of our measurements of $Q_0$ enhancement factor as a function of temperature. The values of note from the plot are the 45K and 77K values which are most relevant for the UCXFEL and C$^3$ use cases. These and select other values are shown in Table \ref{table:values}. We also from the $Q_0$ enhancement compute the surface resistivity values. In addition, to measuring the temperature dependence of the cavity quality factor we also measure the detuning of the cavity itself by tracking the minimum of the cavity S11 value. The detuning when paired with NIST log fitting for cryogenic copper coefficients of thermal expansion can then be used as a secondary verification measurement of the temperature of the cavity. 

To address some of the systematics, the cavity data was recorded on warming of the cavity in order to better ensure thermalization at each temperature. As a note on the temperature dependence of the geometric, our CST simulations show a relatively minute impact on the low level RF figures of merit that were measured and reported here. There is more significant impact on the coupling coefficient which were here note are necessary for data analysis consideration but are omitted simply for brevity.

\section{Discussion}

First of all, the LLRF measurements have immediate implications for the UCXFEL \cite{UCXFEL} and more long term implications for the Cool Copper Collider \cite{nanni2023status}. As shown in the Table \ref{table:values}, the two considered operation points are $45$ and $77$K corresponding to the cryocooled UCXFEL photoinjector and the liquid nitrogen cooled linac sections to be used in both concepts. Quality factor enhancement approximately $4.61\pm 0.05$ and $2.89\pm 0.05$ respectively are necessary numbers for RF design in the the two cases. In the Cband regime to our knowledge these are reported here for the first term for a cavity manufactured to accelerating cavity specifications. 

In addition to the Q factor enhancement, the most significant observable implication is the impact on the RF pulse heating. Our model predicts that there is an optimum working temperature for a cryogenic normal conducting cavity definitively based on minimizing pulse heating. During the initial stages of the commissioning of CYBORG tests RF pulse heating as a a function of temperature are currently underway and will provide an additional measurement in addition to the S11 reflection coefficient \cite{lawler:ipac2023-tupa039}.

Now with respect to the bulk material properties, we should clarify the physical significance of what RRR means. In our consideration we use this as a free parameter derived ex post facto from experiment and incorporated into the Bloch-Gruneisen model. Strictly speaking this FoM is only a proxy for the quantity of interest which is the material purity in terms of defects, grain boundaries etc. We can then refer to high RRR as high purity metal and lower RR as less pure. Less pure here need not refer to a failure of metallurgy, rather it can also refer to intentional alloying. The main implication of the models presented here is that there may exist local surface resistivity minima for purity. More specifically, if one is willing to make the voyage to single digit kelvin, an alloy with purity corresponding to below RRR250 is preferable. 

If higher temperatures are the only option then a more pure metal is required with RRR450 matching the performance at 25K. So hardening the material with alloying CuAg alloys then presents an appealing option to reduce the thermal loads required. In addition, arbitrary purity is known to lead to unmanageable soft copper from the machining perspective \cite{ekin2006experimental} so hardening at room temperature first before cooling may be further be a necessity. We have begun to consider this case with alloys, some of the measurements reported here \cite{lawler:ipac2023-tupa038}. We reiterate this here for the sake of saying that the future testing of the pillbox tests here are not only intending to iteratively reduce temperature to below 30K but also repeat the $R_s$ measurements for varying alloys, especially CuAg, using this pillbox template. 


\begin{table}[h!]
\centering
\begin{tabular}{||c c c c||} 
 \hline
 5.718 GHz cavity  & 295K & 77K & 45K \\ [0.5ex] 
 \hline\hline
 $\Delta f$ [MHz] & 0 & 17.8 &  18.7\\
 $Q_0$ enhancement & 1 & $2.89\pm 0.05$ & $4.61\pm 0.05$ \\ 
 $R_s$ [$\Omega$ m] & $3\times 10^{-2}$  & $1.05\times 10^{-2}$ &  $6.5\times 10^{-3}$\\ [1ex]
 \hline
\end{tabular}
\caption{Measured properties for the two relevant use cases: the 45K operation point for UCXFEL and the 77K operation point for $C^3$}
\label{table:values}
\end{table}


Finally we can further the discussion of the implications beyond the scope of cryogenic cavity development by considering the new thin film model as a possible novel physical environment in which to study thin film physics. We consider the possibility of going beyond the scope of linear accelerator development and instead using cryogenic RF cavities as a new test environment for the study of viscous electron flow. This is an area of intense study for condensed matter and new examples of viscous electron are of certain interest \cite{Polini_2020,doi:10.1143/JPSJ.25.390}. 

\section{Conclusion}

We are here convinced of the validity of our surface physics understanding for a Cband resonant cavity down to at least 38K. With a Q factor enhancement of $4.61\pm 0.05$ at 45K and $2.89\pm 0.05$ at 77K this has immediate impacts on the development of cryogenic copper linear accelerator development, especially UCXFEL and $C^3$. Preparation and machining become the important knobs for precision study but insofar as these are established processes in accelerator physics, we can envision here low level RF cavities as test beds for material physics properties relevant to high gradient cavities in parallel to the CYBORG beamline at UCLA. Future studies will include in order of chronology, a more in depth analysis of $R_s$ in the CYBORG photogun from the standpoint of RF heating, cooling of the cavity to the limits of our cryocooler $<30K$, and the manufacture of equivalent cavities with hard CuAg alloys. We also begin to consider cryogenic RF cavities as possible novel test beds for studying thin film and surface physics in a novel environment. 

\section{Acknowledgements}

This work was supported by the Center for Bright Beams, National Science Foundation Grant No. PHY-1549132 and DOE HEP Grant DE-SC0009914


\printbibliography

\end{document}